\documentclass[iop]{emulateapj}
\usepackage{apjfonts}
\shorttitle{A D3 FLARE}
\shortauthors{LIU ET AL.}
\submitted{Received 2013 April 12; accepted 2013 June 25; published --}
\journalinfo{Accepted to ApJ (06/25/13)}
\usepackage[colorlinks=true,linkcolor=blue,citecolor=blue,urlcolor=blue]{hyperref}

\newcommand{\ha}{H$\alpha$}
\newcommand{\sm}{$\sim$}
\newcommand{\yohkoh}{\textit{Yohkoh}}
\newcommand{\goes}{\textit{GOES}}
\newcommand{\hxrbs}{Hard X-Ray Burst Spectrometer}
\newcommand{\Smm}{\textit{Solar Maximum Mission}}
\newcommand{\smm}{\textit{SMM}}
\newcommand{\xrp}{X-Ray Polychromator}
\newcommand{\kms}{km~s$^{-1}$}

\begin{document}
\title{H\lowercase{e~{\sc i}} D3 OBSERVATION OF THE 1984 MAY 22 M6.3 SOLAR FLARE}
\author{Chang Liu\altaffilmark{1}, Yan Xu\altaffilmark{1}, Na Deng\altaffilmark{1}, Jeongwoo Lee\altaffilmark{1,2}, Jifeng Zhang\altaffilmark{1,3}, Debi Prasad Choudhary\altaffilmark{4}, and Haimin Wang\altaffilmark{1}}
\affil{$^1$~Space Weather Research Laboratory, Center for Solar-Terrestrial Research, New Jersey Institute of Technology,\\University Heights, Newark, NJ 07102-1982, USA; \href{mailto:chang.liu@njit.edu}{chang.liu@njit.edu}}
\affil{$^2$~School of Space Research, Kyung Hee University, Yongin 446-701, Republic of Korea}
\affil{$^3$~Department of Mechanical and Industrial Engineering, New Jersey Institute of Technology, University Heights, Newark, NJ 07102-1982, USA}
\affil{$^4$~Physics and Astronomy Department, California State University Northridge, 18111 Nordhoff Street, Northridge, CA 91330-0001, USA}

\begin{abstract}
He~{\sc i}~D3 line has a unique response to the flare impact on the low solar atmosphere and can be a powerful diagnostic tool for energy transport processes. Using images obtained from the recently digitized films of Big Bear Solar Observatory, we report D3 observation of the M6.3 flare on 1984 May 22, which occurred in an active region with a circular magnetic polarity inversion line (PIL). The impulsive phase of the flare starts with a main elongated source that darkens in D3, inside of which bright emission kernels appear at the time of the initial small peak in hard X-rays (HXRs). These flare cores subsequently evolve into a sharp emission strand lying within the dark halo simultaneously with the main peak in HXRs, reversing the overall source contrast from $-5$\% to 5\%. The radiated energy in D3 during the main peak is estimated to be about 10$^{30}$~ergs, which is comparable to that carried by nonthermal electrons above 20~keV. Afterwards the flare proceeds along the circular PIL in the counterclockwise direction to form a dark circular ribbon in D3, which apparently mirrors the bright ribbons in \ha\ and He~{\sc i} 10830~\AA. All these ribbons last for over one hour in the late gradual phase. We suggest that the present event resembles the so-called black-light flare that is proposed based on continuum images, and that D3 darkening and brightening features herein may be due to, respectively, the thermal conduction heating and the direct precipitation of high-energy electrons.
\end{abstract}

\keywords{Sun: activity  -- Sun: flares -- Sun: magnetic topology -- Sun: X-rays, gamma rays}

\section{INTRODUCTION}\label{sect1}
\ha, the most commonly used line in solar observations, is broad and optically thick, and responds to a wide range of features in the solar chromosphere. In contrast, He~{\sc i}~D3 line (located at 5876~\AA\ near the Na~{\sc i} doublet D1 and D2) is narrow and mostly optically thin, and is sensitive to nonthermal excitation. For the latter reason, the D3 line is considered better suited to the study of higher energy phenomena than \ha. The most unique property of D3 is that it appears in absorption in surges, eruptive filaments, flare ejecta, and weak flares, and turns into emission only in intense flares. Compact and transient bright cores or strands are often observed in D3 during the impulsive phase of major flares, displaying a close spatiotemporal association with footpoint-like white-light (WL) and hard X-ray (HXR) emissions \citep[e.g.,][]{labonte79,zirin80,zirin81,feldman83,tanaka85,zirin90,wang96}. D3 emission can reach about twice the photospheric intensity during flares, and this means that the medium has a high temperature (\sm2$\times10^4$~K) and high density (\sm10$^{13}$~cm$^{-3}$) flaring plasma. Therefore, D3 may provide a diagnostic of the main flare energy source in the low atmosphere complementary to WL and HXRs, as well as a sensitive probe for tracing the flare development. The simultaneity of D3 and HXR footpoint emissions could indicate that accelerated electrons penetrate down to at least the low level as indicated by the above density \citep{zirin88}. \citet{wang96} observed during the initial HXR peak of an M2.7 flare that D3 emission only occurred at one of the HXR footpoints with a harder spectrum. Another X12 flare studied by \citet{tanaka85} started with multiple compact brightenings in D3 and WL, two of which grew to the main flare footpoints when the HXR spectrum hardened with a sharply increasing flux. They further developed into two ribbons in D3 above sunspots around the soft X-ray (SXR) peak, mainly due to the heating from the hot thermal plasma that is conducted down to dense layers. The D3 line has been dubbed ``a gold mine of information'' on flares \citep{zirin88}; nonetheless, it has been largely ignored in recent flare observations.

As one of the strongest He lines in the visible spectrum, the D3 line results from transitions between the $2p\ ^{3}P$ and $3d\ ^{3}D$ terms of the He~{\sc i} triplet (ortho-helium), which involves a metastable level about 20~eV above the ground state. One mechanism to populate the metastable level is that the coronal EUV radiation ionizes neutral He atoms primarily in the singlet ground state at typical chromospheric conditions, and subsequent recombinations produce a larger population of He in the triplet systems hence strengthening the D3 features \citep[e.g.,][]{mauas05,centeno08}. This photoionization-recombination (PR) scenario is supported by the observation of the bright D3 band in a narrow height range of 1000--2000~km above the solar limb, which is however not observed when over coronal hole regions \citep{zirin75}. It is noted that the EUV photons only help populating the triplet levels, which then subsequently radiate by scattering the photospheric emission. Thus in order for D3 to turn into emission against the disk, direct collisional excitation (CE) must exceed the photospheric radiative excitation. It can be deduced from this requirement that at $T \approx 2 \times 10^4$~K, the transition from absorption to emission of D3 in a flare would occur as the density of the heated atmosphere reaches $N \gtrsim 5 \times 10^{12}$ cm$^{-3}$ \citep{zirin88}. Thus far, only a single event exhibiting this implied reversal of D3 intensity has been reported. It was the SOL1978-07-10T17:35\footnote{Following the IAU solar target naming convention \citep{leibacher10}.} M8 flare, in which a dark ring (circular ribbon) in D3 around a satellite spot in the preflare state immediately preceded two bright D3 emission kernels, which supposedly lie below the ring \citep{zirin80}.

Intriguingly, such an evolution from absorption to emission is very reminiscent of the so-called ``negative flares'' on stars, in which a preflare dip in the continuum level, i.e., a depletion of amplitude by \sm20\% with a duration of seconds to minutes, is seen right before the continuum brightening \citep[e.g., ][and references therein]{hawley95}. \citet{henoux90} termed such a phenomenon ``black-light flares'' (BLFs) and proposed that continuum BLFs could also happen on the Sun for some 20~s before WL flares (WLFs). The underlying mechanism could be that nonthermal ionization of hydrogen by the bombardment of electron beams on a cool atmosphere first produces an increase of H$^-$ opacity, and its diminutive effect is overtaken by the enhanced emissivity from a fully heated atmosphere near the flare maximum phase \citep{grinin83,henoux90,ding00,ding03b}. Observations and studying of BLFs would certainly be important for solar and stellar flare physics, as further information can be obtained on the heating of the low atmosphere in the flaring process. Nevertheless, solar BLFs in continuum are extremely rare. \citet{van94} surveyed all nine \yohkoh\ WLFs greater than \goes-class M6 but found no unambiguous evidence for the BLF association. A possible example presented by \citet{henoux90} is a negative intensity contrast up to \sm5\% at 5500~\AA\ continuum a few minutes before the SOL1981-07-26T13:53 X3 WLF. Another candidate is the SOL2001-03-10T04:05 M6.7 WLF observed near the Ca~{\sc ii} 8542~\AA\ line, for which \citet{ding03b} showed a dip of \sm1--2\% in its early phase before the continuum emission.

Nonthermal electron beams not only can lead to BLFs in continuum, but also are able to drastically change spectral lines. Using non-LTE calculations, \citet{ding05} demonstrated that in the nonthermal case, the collisional ionization followed by recombinations (CR) may significantly contribute to populating the He triplet levels corresponding to the He~{\sc i} 10830~\AA\ line, besides the PR and CE mechanisms. Importantly, the influence of CR on the 10830 line, when compared to the case without the nonthermal effects, was modeled by the authors. In their result the nonthermal effect can cause a stronger absorption early in a flare and a stronger emission at the flare maximum. We consider that such a transition from absorption to emission of spectral lines, if observed, could be regarded as a variant of BLFs analogous to those in continuum. In this context, the aforementioned event of a preflare dark ring followed by bright kernels \citep{zirin80} turns out to be a very rare instance of D3 BLFs. To our knowledge, no study has been carried out about the nonthermal effects on the He~{\sc i} D3 line.

In this paper, we present a detailed analysis of the SOL1984-05-22T15:03 M6.3/2B flare, taking advantage of the unique D3 observation made at the Big Bear Solar Observatory (BBSO). The most prominent characteristics of this event include (1) the main D3 flare source first darkens in intensity then changes to emission at the time of the HXR peak, which makes an excellent candidate for BLFs, and (2) in the late phase, there form two dark ribbons in D3 mirroring their bright counterparts in \ha\ as well as He~{\sc i} 10830~\AA. We also compare the radiated energy in D3 with the energy carried by electron beams at the flare peak time, and discuss the event progression using \ha\ images. The plan of the paper is as follows: in Section~\ref{sect2}, we describe the data sets and the reduction procedure. In Section~\ref{sect3}, we present the main results of data analysis and discuss their implication. Major findings are summarized in Section~\ref{sect4}.

\section{OBSERVATIONS AND DATA REDUCTION} \label{sect2}
BBSO has a long tradition of flare observations at the \ha\ wavelength. In addition, BBSO 
started from 1973 to take high-resolution images in the D3 line using its 25 or 65~cm telescopes \citep{feldman83}. All the BBSO data obtained before 1995 were recorded on 35~mm films, a large portion of which have recently been digitized (see \citealt{wang12} for a fuller description). The film data have also been publicly released,\footnote{\url{http://sfd.njit.edu}} which opens the door for studying a wealth of historical solar phenomena. The present event (SOL1984-05-22T15:03 M6.3) is selected from our survey on D3 flare observations.

We mainly used the digitized partial-disk D3 and \ha\ images taken by the east and west benches, respectively, of the 25~cm telescope, with a pixel size of \sm0$\farcs$15 and a cadence of \sm15~s. These images are capable of offering sub-arcsecond resolution when the seeing condition at BBSO is excellent (e.g., the later phase of the flare event under study). A non-linear conversion table supplied by the digitizer, which transforms the original 12 bit digitized data to 8 bit, was applied to deal with the nonlinear intensity response of films. The image alignment was implemented with sub-pixel precision, and intensity was normalized to that outside the flaring region in a quiet-Sun area. A de-stretching algorithm \citep{shine94} using running references was applied in order to reduce the atmospheric distortion. We define the image contrast as $(I-I_0)/I_0$, where $I$ is the intensity of the feature of interest, and $I_0$ is the intensity of the same feature in the preflare state or that of the undisturbed quiet-Sun background. For \ha\ image time series, we computed the intensity-weighted centroid positions of flare ribbons, and used the relative distance of source centroids to infer the progression of the flare \citep[e.g.,][]{liu10}. Additional auxiliary data used in our study include the digitized full-disk (\sm1$\farcs$4 pixel$^{-1}$), 1 minute cadence \ha\ images taken by the National Solar Observatory (NSO) at Sacramento Peak (SP), and the daily full-disk spectroheliogram observation in 10830~\AA\ made by NSO at Kitt Peak (KP) using its 512-channel magnetograph \citep{livingston76}.

The daily full-disk magnetic field structure of the photosphere in the line of sight was also observed by NSO/KP in Fe~{\sc i}~8688~\AA. We registered D3 and \ha\ images with the NSO/KP magnetogram by matching sunspot and plage areas, with an alignment accuracy estimated to $\lesssim$5\arcsec. To reveal the magnetic structure of the flaring region, we resorted to the potential field extrapolation in Cartesian coordinates.

X-ray observations were used to study the temporal evolution of the flare and understand the nature of the D3 emission. SXR and HXR time profiles were obtained with \goes~{\it 5} satellite and \hxrbs~(HXRBS; \citealt{orwig80}) on board the \Smm~(\smm), respectively. We also utilized HXR spectra fitted to all 15 channels of HXRBS covering the energy range of 24--400~keV, which were computed by an automatic spectral fitting routine using a single power-law model \citep{kiplinger95}. The derived spectral parameters allowed us to evaluate the nonthermal thick-target electron power above a certain low-energy cutoff \citep{brown71}. In addition, SXR images were available from \xrp~(XRP; \citealt{acton80}) on board \smm.

\begin{figure}
\epsscale{1.17}
\plotone{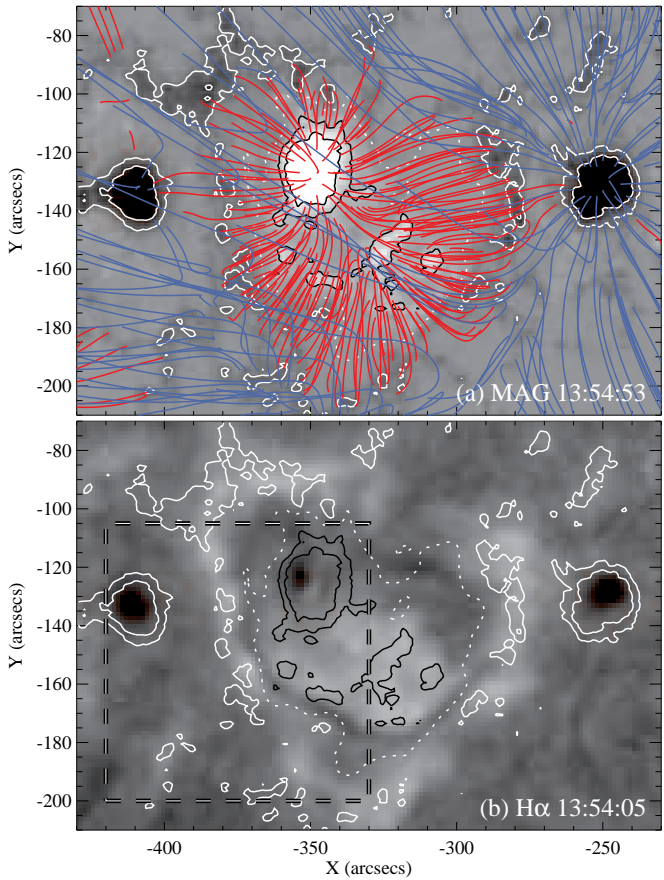}
\caption{Magnetic structure of the flaring region at the leading portion of NOAA AR 4492. (a) A NSO/KP magnetogram (scaled from $-$800 to 800~G) overplotted with the PIL (dotted line), and the closed (red) and open (blue) field lines extrapolated using a potential field approximation. The levels of magnetic field contours are $\pm$300 and $\pm$700~G. (b) A preflare NSO/SP \ha\ image overplotted with the same magnetic contours as well as the PIL. The dashed box indicates the FOV of Figure~\ref{f3}. All the images in this paper are aligned with respect to 1984 May 22 14:50~UT. \vskip 2mm \label{f1}}
\end{figure}

\begin{figure}
\epsscale{1.18}
\plotone{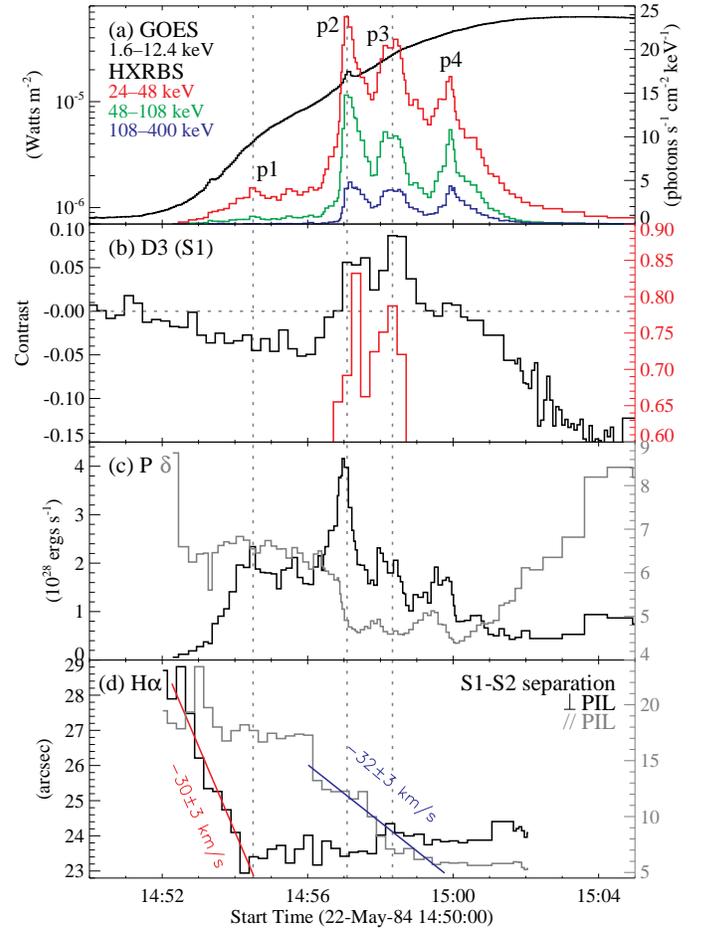}
\caption{Temporal evolution of the flare in multiwavelengths. (a) \goes\ SXR flux overplotted with HXRBS photon fluxes. The 48--108 and 108--400~keV fluxes are times 10 and 30, respectively. (b) Mean intensity contrast (black) of the D3 source S1 (the boxed region in Figure~\ref{f3}(f)) and that of its brightest core region with a contrast enhancement greater than 0.6 (red). (c) Thick-target electron spectral index (gray) and electron power for a low energy cutoff of 20~keV (black). (e) Separation of \ha\ ribbons S1 and S2 in the directions perpendicular and parallel to the PIL. \vskip 2mm \label{f2}}
\end{figure}

\section{RESULTS AND ANALYSIS}\label{sect3}
In this section, we first give a brief account of the magnetic field structure of the flaring region in the preflare state. We then concentrate on describing characteristic flare activities in D3, HXR, as well as \ha\ wavelengths throughout the event. More dynamic detail can be seen in the accompanying animations in the online journal.\footnote{For maximal use of the field of view (FOV), the images in the animations were not corrected for the solar $p$-angle.}

\subsection{Magnetic Field Structure}
According to USAF/NOAA reports, the active region of interest NOAA AR 4492 started to appear in the $\beta\gamma\delta$ magnetic configuration from 1984 May 19 and showed growth till May 21. The AR decayed in size from May 22, when the present M6.3/2B flare occurred in the leading portion of the AR with a $\beta$ configuration (see Figure~\ref{f1}(a)). A prominent property of magnetic field structure of this flaring region is that the central spot of positive polarity seems to be encircled by the surrounding negative field, forming a quasi-circular magnetic polarity inversion line (PIL; the dotted line). This is well delineated by the circle-like \ha\ filament, which lies closely along the PIL (Figure~\ref{f1}(b)). Different from the classical eruptions, this filament does not show a preflare activation and largely survived after the flare, which is clearly seen in the time-lapse \ha\ movies. This feature may indicate that the flare-associated magnetic reconnection could occur above the filament \citep[e.g.,][]{liu07b}.

Remarkably, the result of the potential field extrapolation model (Figure~\ref{f1}(a)) reveals dome-shaped closed fields (red) overarching the filament, and the closed dome is enveloped by open field lines (blue) stemming from the surrounding regions. The overall structure then well portrays a fan-spine configuration, which is usually associated with a coronal null point. It has been found that three-dimensional (3D) magnetic reconnection could occur at such a null-point \citep{lau90,torok09}, especially when pronounced asymmetry is present (as for this AR; see Figure~\ref{f1}(a)) \citep{pariat10}. It is also found that sometimes flares associated with such 3D reconnection may exhibit a central compact kernel situated within an extended circular flare ribbon \citep[e.g.,][]{masson09}. In fact, observational signatures of the present flare bear much resemblance to those described in this scenario, and we will discuss them in more detail in Section~\ref{progress} below.

\begin{figure*}
\epsscale{1.17}
\plotone{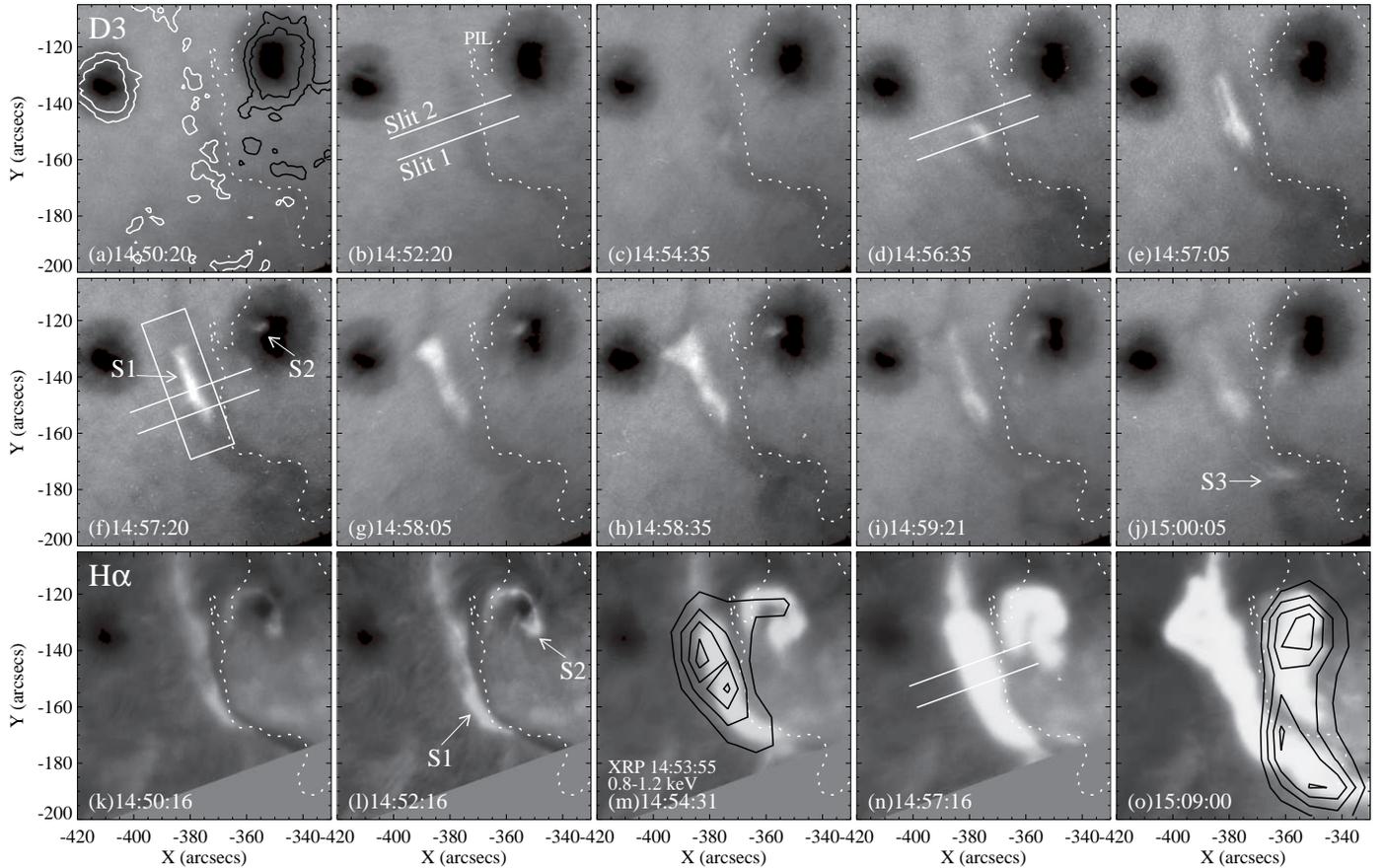}
\caption{Time sequence of D3 ((a)--(j)) and \ha\ ((k)--(o)) images during the flare impulsive phase. The box in (f) marks the area for calculating the overall light curve of the D3 source S1 (black) in Figure~\ref{f2}(b). The white lines in (b), (d), (f), and (n) are slits perpendicular to the flare ribbons/strands, along which the intensity profiles are drawn in Figure~\ref{f4}. Slits 1 and 2 pass through the D3 source maxima in (d) and (f), respectively. The black contours (30\%, 50\%, 70\%, and 90\% of the maximum flux) in (m) and (o) represent XRP SXR images. The magnetic field contours in (a) are the same as those in Figure~\ref{f1}, and the dotted line is the PIL. The image FOV is denoted by the dashed box in Figure~\ref{f1}(b).\\
{\vskip -2mm (Animations of this figure are available in the online journal.)} \vskip 2mm \label{f3}}
\end{figure*}

\subsection{Event Evolution in D3}
\subsubsection{A BLF}
According to \goes\ and HXRBS X-ray fluxes, the event started at \sm14:51~UT and peaked at 15:03~UT on 1984 May 22, and had four conspicuous HXR spikes p1--p4 (see Figure~\ref{f2}(a)). The initiation of the flare in D3 images is featured with a gradual darkening of the elongated main source S1 above the weak positive magnetic field region (cf. Figures~\ref{f3}(a)--(c) and (f)). It is notable that within this darkening source S1, two D3 kernels begin to brighten up from \sm14:54:30~UT (Figures~\ref{f3}(c) and (d)) cotemporal with the first HXR spike p1. The two bright kernels then appear to extend in the direction parallel to the PIL to form an emission strand lying within the dark halo-like structure at 14:57:05~UT (Figure~\ref{f3}(e)), which is simultaneous with the most intense spike p2 in HXRs from 24--400~keV. An intriguing observation is that the strand apparently reaches its highest brightness a little later at 14:57:20~UT (Figure~\ref{f3}(f)), at the time of a slightly harder HXR spectrum (Figure~\ref{f2}(c)). From \sm14:57:50~UT, S1 broadens and enhances particularly in its northern end, and during the time period of the HXR spike p3, it spreads eastward into the sunspot region (Figures~\ref{f3}(g)--(i)).

To more quantitatively assess the temporal relationship between D3 emission and flare energy release, we measure the mean contrast of the D3 source S1 in a boxed region (see Figure~\ref{f3}(f)) and draw its time profile in Figure~\ref{f2}(b) (black line). It can be unambiguously seen that S1 first darkens to develop an intensity dip, reaching a contrast amplitude of $-$5\% in about four minutes from 14:52 to 14:56~UT despite of the brightening kernels. It then quickly turns to positive contrast after being overtaken by the bright emission strand, and peaks with an enhancement of \sm5--10\% concurrently with the HXR spikes p2 and p3 with hardened spectra. It is worth mentioning that the highest D3 intensity of S1 actually occurs at the time of the strongest HXR peak p2 as expected and shows a contrast up to 100\% as previously observed in other flares of similar magnitude \citep[e.g.,][]{zirin80}. This is clear in the time profile of averaged contrast with only the contrast greater than 60\% (red line in Figure~\ref{f2}(b)). Here the most significant result is the intensity reversal of S1 from absorption to emission, which qualifies this event as a rarely observed BLF, and for the first time directly evidences the unique property of BLFs in the D3 line.

We note that the absorption feature in D3 was generally interpreted as due to heating of the dense layer by thermal conduction \citep{zirin88}. This could explain the darkening of S1 below the BLF level in the later thermal phase (after \sm15:02 UT). More importantly, D3 and HXR emissions in the earlier impulsive phase are simultaneous with good correlation, which implies electron beam heating as suggested before \citep{feldman83,zirin88}. In other words, D3 emitting sources in the flare impulsive phase could indicate the location of electron precipitation. As supporting evidence, a compact and weak D3 bright kernel S2 is located at the umbral region of the western sunspot with negative polarity. It becomes barely visible since \sm14:56~UT, peaks at the HXR spike p1 around 14:57~UT with a contrast of \sm70\% against the dark sunspot, and propagates southward after \sm14:58~UT (Figures~\ref{f3}(d)--(i)). We speculate that S1 and S2 are a pair of conjugate sources at the feet of reconnecting flare loops, which are indicated by the loop-like structure in SXRs crossing over the PIL (contours in Figure~\ref{f3}(m)). As it lies within the stronger umbral field region, the compactness of the source S2 in D3 could be due to the magnetic mirroring effect \citep[e.g.,][]{kundu95}. It is also noticed that the last HXR spike p4 at 15:00~UT is reflected in D3 as a newly formed source S3 in the south (Figure~\ref{f3}(j)). Flaring loops could connect S2 and S3 at this stage, as imaged in SXRs at a later time (Figure~\ref{f3}(o)).

\subsubsection{Bright Core-Dark Halo Structure}
\begin{figure}
\epsscale{1.17}
\plotone{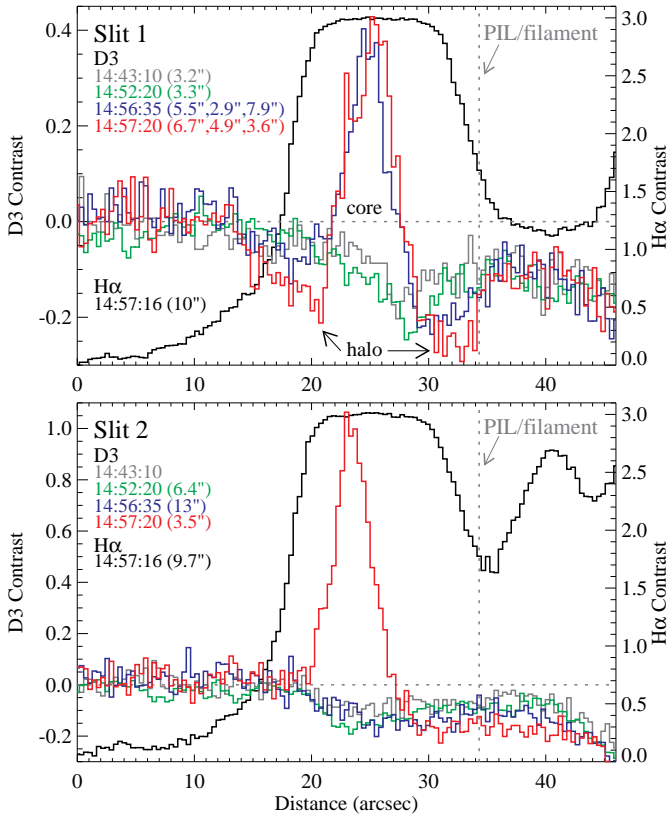}
\caption{D3 and \ha\ intensity profiles along the slits 1 and 2 (as drawn in Figure~\ref{f3}) at different times. The distance is measured from the southeastern ends of the slits. Gaussian fits are employed to derive the characteristic sizes (FWHM values are given in brackets) and peak positions of the dark halo and/or bright core emissions. Note that the central \ha\ emission is saturated. \label{f4}}
\end{figure}

We place two slits perpendicular to the elongated source S1 as drawn in Figure~\ref{f3}, and present the intensity profiles along them in Figure~\ref{f4}, in order to more closely examine the dark absorption and bright emission features in D3 and their temporal evolution. Gaussian fits are conducted to estimate the characteristic sizes and peak positions.

The slit 1 crosses the maximum of the D3 kernel in Figure~\ref{f3}(d). The intensity profile along it already displays a discernible darkening about 8 minutes before the flare (gray line), possibly implying the occurrence of the preflare heating. After the flare onset at 14:52:20~UT, the darkening contrast deepens and shifts toward the PIL (green line). The later bright emission kernel at 14:56:35 (blue line) and 14:57:20~UT (red line) peaks \sm4\arcsec\ to the east. However, (1) the darkening contrasts on either sides of the emission peak both deepen consecutively with a step of \sm0.05\%. (2) The western darkening keeps moving toward the PIL by \sm4\arcsec. (3) While the eastern darkening broadens, the western darkening narrows at the HXR peak time (red line) to reach the largest contrast of about $-$3\%. It is reasonable to surmise that the preference of D3 darkenings toward the PIL is caused by the enhanced density near the filament lying along the PIL region (see Figure~\ref{f3}(k)), which provides a favorable condition for the D3 line formation. These results thus clearly illustrate a D3 flare source structure comprising a low-lying bright emission core (due to nonthermal beam heating) encompassed by darkening halos (due to thermal conduction heating). Consistent with this picture, the \ha\ intensity profile along the slit appears, in contrast, as a broad bump (the central part is saturated) with about twice the full-width at half-maximum (FWHM) width, straddling the D3 bright core and dark halos. It is also appealing to suggest that the bright core-dark halo source structure in D3 is in analogy with the core-halo morphology of WLF sources in the photospheric level \citep{xu06}, where both the core and halo are bright and might be produced by direct electron heating and chromospheric back-warming, respectively.

The slit 2 crosses the maximum of the D3 strand S1 around the HXR peak time (Figure~\ref{f3}(f)). Its central region first darkens (green line) then reverts back (blue and red lines) to become a much intense emission core peaking at 100\% contrast while with a similar FWHM width compared to the case of the slit 1. It does not seem that a dark halo developed in the east of the core; nevertheless, a darkening region is similarly built up in the western region close to the PIL/filament.

\subsubsection{Flare Energy Release}
We further choose the flare peak p2 in D3 and HXRs, which is well defined both temporally and spatially, to roughly evaluate the energy budget. The underlying assumption is that the D3 line emission could be optically thick in flares \citep{svestka76,zirin88}, hence the gas radiates with a source function close to the Planck function for the corresponding temperature. In the ``canonic'' flare model \citep[e.g.,][]{wang98c}, the intensity contrast enhancement of a flare can be given by

\begin{equation}
\frac{\Delta I}{I}=\frac{B(T+\Delta T, \lambda)-B(T, \lambda)}{B(T, \lambda)} \ ,
\end{equation}

\noindent where $B$ is the Planck function and $\Delta T$ is the temperature perturbation due to flare heating. The flare radiative loss flux can then be written as

\begin{equation}
\Delta L=4\sigma T^3 \Delta T \ \rm erg \ cm^{-2} \ s^{-1} \ ,
\end{equation}

\noindent where $\sigma$ is the Stefan-Boltzmann constant \citep{najita70}. For $n \approx 10^{13}$~cm$^{-3}$ that is required for D3 emission, $T$ is about 6400~K according to the VAL-F atmospheric model \citep{vernazza81}. We use this model because it was constructed for the magnetic network regions suitable for the present flare, hence is better than other models for the quiet-Sun conditions. If we integrate $\Delta L$ over the source S1 and take a duration of 30~s, the radiative energy in D3 during the peak p2 is estimated to be about 1.3~$\times$~10$^{30}$~ergs. This is comparable to the nonthermal energy $E_{\rm nonth} \approx 1.2 \times 10 ^{30}$~ergs released by the thick-target electrons greater than 20 keV around the peak p2 (Figure~\ref{f2}(c)). Obviously, such a comparison is meaningful provided that electrons with initial energies of 20~keV and above can penetrate to densities \sm10$^{13}$~cm$^{-3}$. A simplified equation relating the observed photon energy $\epsilon$ to the column density $N$ of a certain layer, into which the parent electrons of energy $E$ can precipitate, is

\begin{equation}
N=10^{20} \ {\rm cm}^{-2} \ (\frac{\epsilon}{20 \ {\rm keV}})^2
\end{equation}

\noindent \citep{brown02}. Since $E \approx \epsilon$ holds for a steep electron spectrum ($\delta \approx 5$ around the peak p2), the above condition can be shown to be met for a scale height of \sm200~km. Electrons with lower energies may be stopped higher up and produce emissions in other lines (e.g., \ha). Therefore, the similarity between the amount of energy released in the D3 emission and that carried by nonthermal electrons bolsters the inference that the D3 line emission during the impulsive phase is able to reveal the electron precipitation sites, with an advantage of density diagnostic compared to HXRs.  

\begin{figure*}
\epsscale{1.11}
\plotone{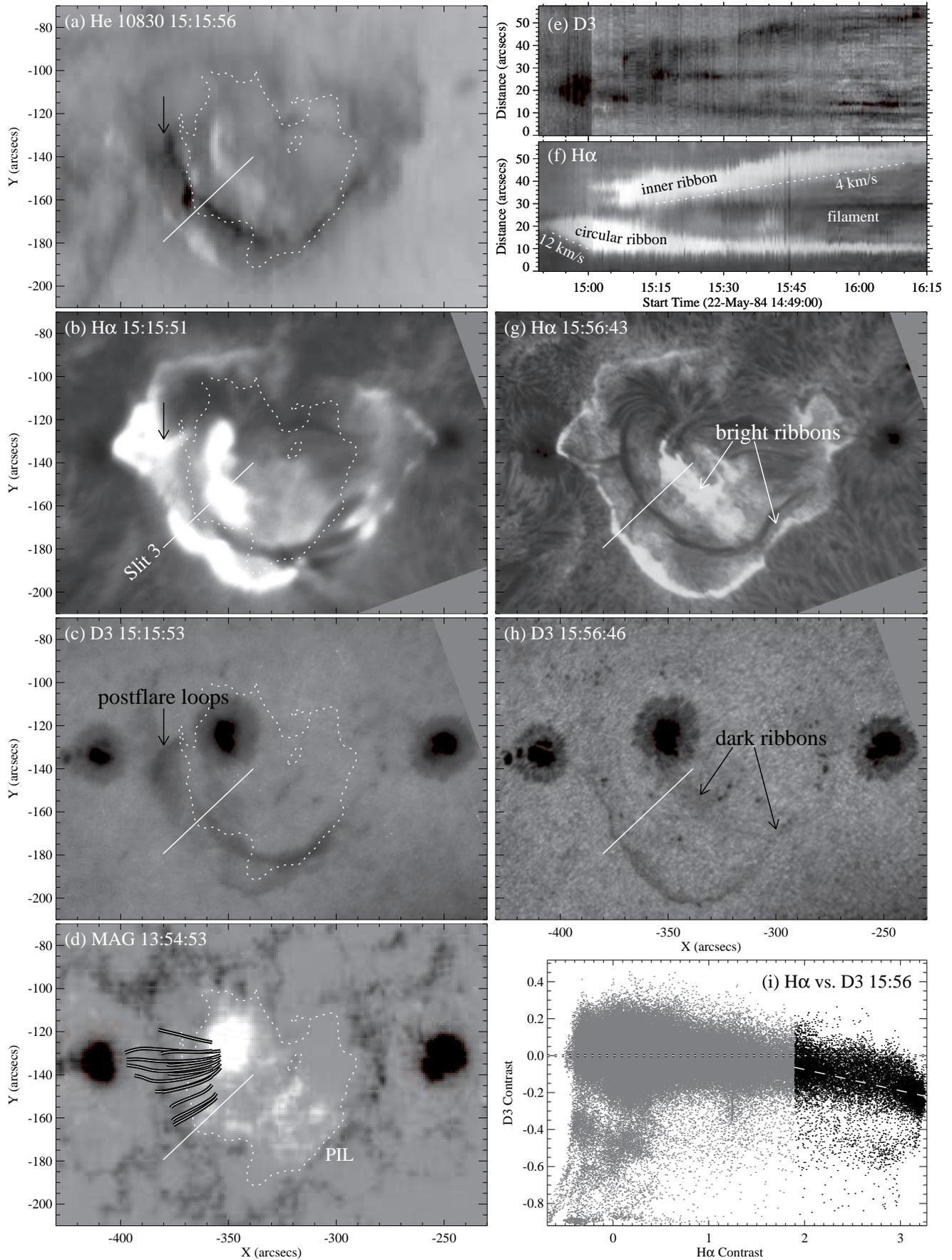}
\caption{Dark circular ribbons in D3 ((c) and (h)) in comparison with their bright counterparts in \ha~((b) and (g)) and He~{\sc i} 10830~\AA~(a), in the late thermal phase under moderate ((b) and (c)) and excellent ((g) and (h)) seeing conditions. The thick lines in (d) show some of the modeled potential field lines. (e) and (f) are time slices using D3 and \ha\ images, respectively, for the slit 3 (white line). The distance is measured from the southeastern end. (i) shows the scatter plot of the \ha\ contrast in (h) versus the the D3 contrast in (g). The white dashed line is a linear fit to the black data points, which correspond to the regions of flare ribbons. \label{f5}}
\end{figure*}

\subsection{Event Evolution in \ha} \label{progress}
It is also instructive to study the progression of the flare emission in \ha. We measure the distance between the centroids of S1 and S2 and plot its temporal evolution in Figure~\ref{f2}(d). Before the HXR peak p1, S1 grows along the PIL while S2 shows a counterclockwise motion toward the PIL (see Figures~\ref{f3}(k)--(m) and the time lapse movies), which mainly leads to a decrease of the centroid distance in the direction perpendicular to the PIL. As S1 and S2 further extend along the PIL northward and southward, respectively, the centroid distance again decreases in the direction parallel to the PIL during the HXR peaks p2 and p3. These non-standard motions of \ha\ ribbons distinct from the classical separation motion away from the PIL have also been observed before in flares involving 3D reconnection in a fan-spine topology \citep{wang12}.

Other supporting evidence includes that (1) after \sm15:00~UT, the event predominantly proceeds along the PIL to form a circular outer ribbon surrounding the elongated but more compact inner ribbon; (2) there occurred around the HXR peak p2 a remote brightening in the northern plage region\sm200\arcsec\ away (see the full-disk \ha\ movie), which has the same positive magnetic polarity as the central spot of the flaring region; and (3) a jet-like eruption of dense material from \sm15:08--15:18~UT might conform to related simulations \citep{pariat10}. We tentatively associate the above flare ribbons/brightenings with the chromospheric mapping of the fan and spine field lines stemming from the coronal null point, as previously done for other circular ribbon flares \citep{masson09,reid12,wang12,deng13}. A full investigation in this context, however, is worth pursuing in a separate study.

\begin{figure}
\epsscale{1.13}
\plotone{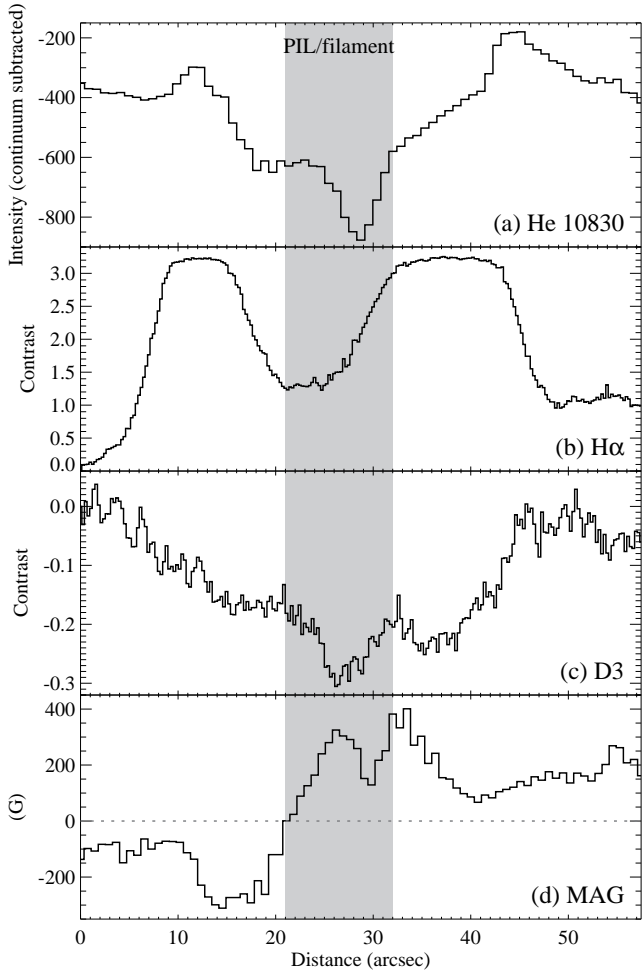}
\caption{Spatial distribution of multiwavelength intensities and magnetic field strength at 15:15 UT along the slit 3 (as drawn in Figures~\ref{f5}(a)--(d)). The distance is measured from the southeastern end. The He~{\sc i} 10830~\AA\ spectroheliograms of NSO/KP are obtained by subtracting the neighboring continuum intensity from the line center intensity at each pixel. The gray region indicates the location of the PIL/filament. \vskip 4mm \label{f6}}
\end{figure}

\subsection{Dark and Bight Circular Ribbons}
This flare is a long-duration event as after the nonthermal impulsive phase, the \goes\ SXR flux decayed to the preflare level in about two hours till around 17~UT. The entire evolution of the event is clearly delineated by the distance-time profiles in D3 and \ha\ along the slit 3 (Figures~\ref{f5}(e)(f)), which lies perpendicular to the PIL and both the outer circular and inner compact ribbons (the white line in Figure~\ref{f5}). The profiles along the slit in multiwavelengths at a specific instance are also drawn in Figure~\ref{f6}. It is obvious that after expanding across the network field during the impulsive phase (cf. Figures~\ref{f5}(f) and \ref{f6}(d)), the outer ribbon stops and remains almost fixed. The inner ribbon, however, keeps propagating northwestward during the late gradual phase at an average speed of \sm4~\kms\ (also see movies). Strikingly, identical dynamics are also exhibited by the dark ribbons in D3 (Figure~\ref{f5}(e)). The continuous ribbon motion leads us to conjecture that there might be a persistent, moderate magnetic reconnection providing continuous injection and deposition of energy into the low atmosphere in the flare late phase. In such a stage, the energy transport mechanism is most probably the thermal conduction \citep[e.g.,][]{czaykowska01}.

Around 15:16~UT, NSO/KP obtained a spectroheliogram in He~{\sc i} 10830~\AA. This provides an opportunity to compare flare signatures at different heights spanning from the bottom to the upper chromosphere using D3, \ha, and 10830~\AA\ images (Figures~\ref{f5}(a)--(c)). Several features are described as follows. (1) The circular filament especially its southern portion appears dark in all wavelengths, which evidences heating of the dense filament materials. The jet-like disruptive activity of the southern segment around this time may be associated with the enhanced darkening in D3. (2) Postflare loops as pointed to by the arrow are dark in D3 and 10830~\AA\ while bright in \ha, and are also reflected by the extrapolation model (thick lines in Figure~\ref{f5}(d)). This may indicate plasma in the postflare arcades at a temperature of \sm2$\times 10^4$~K and a density of 10$^{10}$~cm$^{-3}$ \citep{zirin80}. (3) Most interestingly, the double flare ribbons are bright in \ha\ and 10830~\AA\ while dark in D3, which is distinctly seen as the good correspondence between the peaks in \ha/10830 and valleys in D3 of their profiles along the slit 3 (Figures~\ref{f6}(a)--(c)). Since both 10830 and D3 lines are produced by ortho-helium, this seems to imply that the 10830 line is more easily excited than D3, and becomes emission even by thermal conduction.

We explore further the dark D3 and bright \ha\ circular ribbons around 16~UT when the seeing condition turned excellent. It is evident that the dark ribbons in D3 demonstrate a nearly exact uniformity with the bright ribbons in \ha\ (cf. Figures~\ref{f5}(g) and (h)). Such a correlation shows up as a negative trend of ribbon pixels as fitted by the white-dashed line in the scatter plot of \ha/D3 contrasts in Figure~\ref{f5}(i). Since the \ha\ emission is largely not saturated at this time, it points to a \sm20\% D3 darkening at a \sm300\% \ha\ enhancement level, consistent with the results in Figure~\ref{f4}. Dark flare ribbons in D3 are rarely observed. As a comparison, the aforementioned dark D3 ring observed by \citet{zirin80} has a much smaller (\sm$\frac{1}{7}$) spatial scale and exists for only a short period of time (\sm5 min) in the early stage of a surge. The author interpreted the dark circular D3 ribbon as the projection of a shell of dense (\sm10$^{10}$~cm$^{-3}$) and hot (\sm2$\times$10$^4$~K) gas. Noticeably, this vision of the shell is well in line with the dome-like fan surface as we speculate for this event.

\section{SUMMARY AND DISCUSSION}\label{sect4}
In this paper, we have presented a rare observation of a major flare in the He~{\sc i} D3 wavelength, taking advantage of the recently digitized, high spatiotemporal resolution BBSO film data. We also compared the radiative energy in D3 at the flare peak time with the nonthermal electron energy, and discussed the event evolution with the aid of \ha\ images and the potential field extrapolation model. Our results can be summarized as follows.

\begin{enumerate}

\item The flare initiation is characterized by the development of an elongated darkening source S1 in D3, which reaches a negative contrast of \sm5\% in about four minutes (Figures~\ref{f2} and \ref{f3}). Within this dark source, emission kernels are seen to brighten with the first minor HXR peak. Then, simultaneous with the main HXR peak p2 they evolve into a narrow bright strand lying within the dark halo, reverting its overall contrast to an enhancement of \sm5\%. This intensity reversal of S1 in D3 establishes this event as a rare BLF, which is highly reminiscent of those proposed in continuum \citep{henoux90}.

\item The main D3 source S1 displays a bright core surrounded by dark halos (Figures~\ref{f3} and \ref{f4}). The bright emission core and dark absorption halos are suggested to be caused by nonthermal electron heating and thermal conduction heating, respectively. This bright core-dark halo structure might be a unique characteristic of flare sources in D3, and is in analogy to the bright core-halo morphology of WLFs \citep{xu06}.

\item The flare radiative energy loss in D3 during the \sm30~s period of the main HXR peak is about 10$^{30}$~ergs, which is comparable to the deposited energy of nonthermal electrons with initial energies higher than 20~keV. Together with the good correlation between the D3 and HXR time profiles (Figure~\ref{f2}), these strongly suggest that D3 emission is an excellent tracer of electron precipitation in the low atmosphere during the flare impulsive phase.

\item The flare disturbance generally propagates sequentially along the PIL, and ends up with a dark circular ribbon in D3 mirroring the bright ribbons in \ha\ and He 10830~\AA\ (Figures~\ref{f3} and \ref{f5}). The dark D3 ribbons in the late gradual phase well manifests the thermal conduction heating. The non-standard motions of flare ribbons and the dome-shaped magnetic structure of the flaring region (Figure~\ref{f1}) strongly suggest the 3D magnetic reconnection in a fan-spine topology \citep[e.g.,][]{wang12}.

\end{enumerate}

The mechanism for the solar continuum BLFs is different from that for the solar BLFs observed at He~{\sc i} D3 and 10830~\AA. In the former mechanism, the low atmosphere first responds to the precipitation of flare electrons with increased H$^-$ opacity for a few ten seconds at the start of the main HXR burst \citep{henoux90}. In the latter mechanism, the He line turns from absorption into emission when He triplet levels are populated due to CE (for D3; \citealt{zirin88}) and additional CR (for 10830~\AA; \citealt{ding05}) processes. Unlike continuum BLFs that are proposed to occur before WLFs, the present D3 BLF coincides with the first small HXR peak possibly in the precursor stage before the main HXR bursts. To our knowledge, solar BLFs in continuum and 10830~\AA\ have not yet been unambiguously observed.

A comprehensive study of the current event is unavoidably hampered by the limited quality of the aging films. We recall that flare observation in the D3 wavelength had been recognized as an important probe of the flare heating mechanism. A full understanding of the D3 signatures may also help resolve the puzzle of WLF emissions in the deep photosphere \citep{zirin88}. Therefore, new D3 flare observations using the modern instruments and further modeling efforts especially on the nonthermal effects on the D3 line are highly desired, and will shed new light on the flare impact on the low atmosphere.

\acknowledgments
This study is dedicated to Professor Harold Zirin, the founder of Big Bear Solar Observatory, who passed away on 2012 January 3. We are indebted to BBSO staffs for tremendous efforts in obtaining D3 and \ha\ observations on films. We thank the referee for helpful comments. Digitization of BBSO as well as NSO solar film images are carried out by the Space Weather Research Lab of NJIT, with support from NSF grant AGS 0849453 and NASA grant NNX11AC05G. The LOS magnetogram and He 10830 spectroheliogram used here were produced by NSO/Kitt Peak. X-ray data were obtained by \smm, a NASA mission managed by GSFC, and by \goes, a NOAA/NASA project. C.L., Y.X., N.D., J.Z., and H.W. were also supported by NASA under grants NNX13AF76G, NNX13AG13G, and NNX11AO70G, and by NSF under grants AGS 1153424 and AGS 0839216. J.L. was supported by the international scholarship of Kyung Hee University and NASA grant NNX11AB49G. D.P.C was partially supported by NSF grant AGS 0548260.

\end{document}